\documentclass[aps,preprintnumbers,,nofootinbib,superscriptaddress]{revtex4}
\usepackage{times}
\usepackage{amsmath}
\usepackage{amssymb}
\usepackage{graphicx}
\usepackage{subfigure,accents}
\newcommand{\lc}[1]{\accentset{\circ}{#1}}
\usepackage{booktabs}
\usepackage{rotating}
\usepackage{longtable}
\usepackage{bm}
\usepackage[pagebackref=false,colorlinks=,linkcolor=blue,citecolor=blue]{hyperref}
\usepackage{epstopdf}
 \pdfoutput=1
 \usepackage[utf8]{inputenc}
\usepackage{cases}
\usepackage{amsmath,float}
\usepackage{amssymb}
\usepackage{amsfonts}
\usepackage{amssymb}
\usepackage{dcolumn}
\usepackage{bm}
\usepackage{bbm}
\usepackage{graphicx}
\usepackage{xcolor}
\usepackage{array}
\usepackage{subfigure}
\usepackage{hyperref}
\usepackage{wasysym}

\newcommand{\be}{\begin{equation}}
\newcommand{\ee}{\end{equation}}
\newcommand{\ba}{\begin{eqnarray}}
\newcommand{\ea}{\end{eqnarray}}

\newcommand{\gsim}{\mathrel{\hbox{\rlap{\lower.55ex \hbox {$\sim$}}
                   \kern-.3em \raise.4ex \hbox{$>$}}}}
\newcommand{\lsim}{\mathrel{\hbox{\rlap{\lower.55ex \hbox {$\sim$}}
                   \kern-.3em \raise.4ex \hbox{$<$}}}}

\hypersetup{colorlinks=true,
            breaklinks=true,
            pdfstartview=Fit,
            linkcolor=blue,
            citecolor=blue,
            urlcolor=blue}

\newcommand{\lb}{\label}

\newcommand{\bw}{\begin{widetext}}
\newcommand{\ew}{\end{widetext}}

\def\tt#1{\texttt{#1}}

\def\ber{\begin{eqnarray}}
\def\eer{\end{eqnarray}}
\def\beq{\begin{equation}}
\def\eeq{\end{equation}}

\def\tt#1{\texttt{#1}}

\newcommand{\udt}[3]{#1^{#2}_{\phantom{#2}#3}}

\newcommand{\dut}[3]{#1_{#2}^{\phantom{#2}#3}}

\begin{document}

\title{\large\bf Testing Born-Infeld $f(T)$ teleparallel gravity through Sgr A$^\star$ observations}
\author{Kimet Jusufi}
\email{kimet.jusufi@unite.edu.mk}
\affiliation{Physics Department, State University of Tetovo, Ilinden Street nn, 
1200,
Tetovo, North Macedonia}

\author{Salvatore Capozziello}
\email{capozziello@unina.it}
\affiliation{Universit\`a degli studi di Napoli Federico II, Dipartimento di Fisica Ettore Pancini, Complesso
Universitario di Monte S. Angelo, Via Cinthia Edificio 6, 80126 Naples, Italy,\\
Scuola Superiore Meridionale, Largo S. Marcellino 10, 80138 Naples, Italy,\\
Istituto Nazionale di Fisica Nucleare, Sezione di Napoli, Complesso Universitario di Monte S.
Angelo, Via Cintia Edificio 6, 80126 Naples, Italy.}

\author{Sebastian Bahamonde}
\email{bahamonde.s.aa@m.titech.ac.jp}
\affiliation{Department of Physics, Tokyo Institute of Technology 1-12-1 Ookayama, Meguro-ku, Tokyo 152-8551, Japan. and
Laboratory of Theoretical Physics, Institute of Physics,
University of Tartu, W. Ostwaldi 1, 50411 Tartu, Estonia.}

\author{Mubasher Jamil}
\email{mjamil@sns.nust.edu.pk}
\affiliation{School of Natural Sciences, National University of Sciences and Technology, Islamabad 44000, Pakistan}
 
\begin{abstract}
We use observational data from the S2 star orbiting around the  Galactic Center to  constrain  a black hole solution of extended teleparallel gravity models.   Subsequently, we construct the shadow images of  Sgr A$^{\star}$ black hole. In particular, we constrain the parameter $\alpha=1/\lambda$ which appears in the Born-Infeld $f(T)$ model. 
In the strong gravity regime we find that the shadow radius increases with the increase of the parameter $\alpha$. 
Specifically, from the S2 star observations, we find  within 1$\sigma$ that the parameter must lie between $0 \leq \alpha/M^2 \leq 6 \times 10^{-4}$. Consequently, we used the best fit parameters to  model the shadow images of Sgr A$^{\star}$ black hole and then using the Gauss-Bonnet theorem we analysed the deflection angle for leading order expansions of the parameter $\alpha$. It is found that within the parameter range, these observables are very close to the Schwarzschild case.
Furthermore, using the best fit parameters for the Born-Infeld $f(T)$ model we show that the angular diameter is consistent with recent observations for the Sgr A$^{\star}$ with angular diameter  $(51.8 \pm 2.3) \mu$arcsec and difficult to be distinguished from the GR. For the deflection angle of light, in leading order terms, we find that the deflection angle expressed in the ADM mass coincides with the GR, but the ADM mass in the   Born-Infeld $f(T)$ gravity increases with the increase of $\alpha$ and the overall deflection angle is expected to me greater in $f(T)$ gravity. As a consequence of this fact, we have shown that the electromagnetic intensity observed in shadow images is smaller compared to GR. 
\end{abstract}

\maketitle

\section{Introduction}
Black holes  can be currently considered the leading astrophysical laboratories for testing  
General Relativity (GR) as well as any theory of modified and quantum gravity.  They allow tests in strong field regime and potentially are one of the main tools to discriminate among concurring theories of gravity. In 
particular, recent advances 
in optical, radio, X-ray and gravitational wave astronomy 
\cite{Abbott:2016blz,EventHorizonTelescope:2019dse,
EventHorizonTelescope:2019uob,EHT2022-1,EHT2022-2,EHT2022-3} have directly
confirmed the presence of supermassive  black holes in the galactic centers of 
giant elliptical and spiral galaxies, as well as small astrophysical black 
holes of medium and stellar sizes. Due to the  observations of the first     
radio images of the supermassive black hole   at the center of the   
M87* galaxy, by  Event Horizon Telescope  (EHT) collaboration, black hole  
shadows have become a very useful tool to test GR and examine 
whether possible deviations from it are possible 
\cite{CANTATA:2021ktz}.
In such studies,  one first  calculates the
shadows of various black hole
solutions \cite{Shaikh:2019fpu,Wei:2019pjf,Moffat:2019uxp,Firouzjaee:2019aij,
Banerjee:2019cjk,Long:2019nox,Zhu:2019ura,Konoplya:2019goy,Contreras:2019cmf,
Li:2020drn,Kumar:2020pol,Pantig:2020uhp,Xavier:2020egv,Guo:2020zmf,Roy:2020dyy,
Jin:2020emq,Islam:2020xmy,Chen:2020aix,Jusufi:2020zln}
and then confronts them with the  M87* data  
\cite{Davoudiasl:2019nlo,Bar:2019pnz,Jusufi:2019nrn,Sau:2020xau,Belhaj:2020rdb,Kumar:2020yem,Zeng:2020vsj,
Saurabh:2020zqg,Bambi:2019tjh,Vagnozzi:2019apd,Haroon:2019new,Shaikh:2019hbm,
Cunha:2019ikd,Banerjee:2019nnj,Feng:2019zzn,Li:2019lsm,Allahyari:2019jqz,
Rummel:2019ads,Vagnozzi:2020quf,Chang:2020lmg,Kruglov:2020tes,
Ghosh:2020tdu,Psaltis:2020lvx,Hu:2020usx,Li:2021mzq, Addazi, Felix1, Felix2,Felix3}.

GR predicts that the unique astrophysically viable asymptotically flat black hole solution is only described by its mass, angular momentum and charge. In astrophysics, the charge is usually very small so it does not play any role. Further, if one only takes spherical symmetry, the solution is just described by its mass and, thus, the Schwarzschild solution. If one modifies GR, then, there could be more asymptotically flat spherically symmetric black hole solutions. There are several ways of modifying GR \cite{Capozziello:2011et, Faraoni}. One interesting approach is the so-called teleparallel gravity when curvature is zero but torsion is non-zero~\cite{Bahamonde:2021gfp,Aldrovandi:2013wha,Cai,Krssak:2018ywd} (still the compatibility condition holds $\lc{\nabla}_\alpha g_{\mu\nu}=0$). It turns out that there is a theory which is equivalent to GR (having the same predictions and classical field equations), which is known as the teleparallel equivalent of GR (see~\cite{Aldrovandi:2013wha}). The action of this theory is built from the torsion scalar $T$ which is constructed by contraction of torsion tensor with some particular coefficients in front. This scalar $T$ differs by a boundary term with respect to the Levi-Civita Ricci scalar $\mathring{R}$. If one starts from this theory, one can modify its action and formulate modified teleparallel theories of gravity (see~\cite{Bahamonde:2021gfp} for a comprehensive review about them). For example, one can formulate modified theories by having the three scalars constructed by the torsion tensor~\cite{Hayashi:1979qx,Bahamonde:2017wwk}, or by introducing scalar fields, which are the so-called scalar-torsion theories~\cite{Geng:2011aj,Gonzalez:2015sha,Bahamonde:2019shr,Hohmann:2018ijr,Hohmann:2018vle,Hohmann:2018dqh,Hohmann:2018rwf}, or by including the boundary term $B$~\cite{Bahamonde:2015zma,Bahamonde:2019jkf,Bahamonde:2020bbc,Escamilla-Rivera:2019ulu,Caruana:2020szx}, or even by considering other scalars such as the teleparallel Gauss-Bonnet~\cite{Kofinas:2014owa,Bahamonde:2016kba, Capozziello:2016eaz}. Due to the nature of the torsion tensor, it is much simpler than in the Riemannian case to construct theories with second order field equations. The covariant formulation of these theories were already developed in~\cite{Krssak:2015oua,Golovnev:2017dox,Krssak:2018ywd}.

One famous modification in teleparallel gravity is known as $f(T)$ that it is achieved by upgrading the torsion scalar to any arbitrary function~\cite{Ferraro:2008ey,Ferraro:2006jd,Bengochea:2008gz,Krssak:2015oua}. This theory is usually formulated with tetrads and a spin connection which is always pure gauge. This means that it is always possible to choose a specific gauge such that the spin connection vanishes. Further, this theory contains symmetric and antisymmetric field equations. If one imposes that both the metric and the connection respects spherical symmetry, then the two most general tetrads (in the Weitzenb\"ock gauge)~\cite{Hohmann:2019nat} that solve the antisymmetric field equations for $f(T)$ were obtained in~\cite{Bahamonde:2021srr}. The first one is real and so far, there are not non-trivial exact black hole solutions (only perturbed ones)~\cite{Bahamonde:2019zea,Ruggiero:2015oka,DeBenedictis:2016aze,Bahamonde:2020vpb,Golovnev:2021htv,Pfeifer:2021njm}. Further, using numerical and a dynamical approach, it has been found that the real tetrad contains a regular black hole solution in $f(T)$ Born-Infeld gravity~\cite{Bohmer:2020jrh,Bohmer:2019vff}. On the other hand, it has been recently found that for the complex tetrad, there are two exact asymptotically flat black hole solutions for two different theories, a Born-Infeld $f(T)$ and a more complicated one given by non-trivial modifications of GR~\cite{Bahamonde:2021srr}. These two exact black solutions represent the first non-trivial ones in the literature of teleparallel gravity. In the context of scalar-torsion theories, recently, new exact scalarized black hole solutions were obtained in~\cite{Bahamonde:2022lvh}. 

Our aim is to study the astrophysical properties of the exact black hole solution found in Born-Infeld $f(T)$ gravity and use data to constrain its parameter. This particular theory also has the interesting feature that it can describe inflation without introducing a scalar field. This paper is organized as follows. In Section~\ref{review}, we briefly review teleparallel gravity and exact  Born-Infeld  solution  reported recently in  \cite{Bahamonde:2021srr}. In Section~\ref{secsha}, we study the shadow images in  Born-Infeld $f(T)$ gravity. In Section~\ref{secobv}, we constrain the parameter of the solutions using the motion of the S2 star orbit. In Section~\ref{def}, we investigate the deflection angle of light in the weak gravity regime. Finally in Section~\ref{seccon}, we comment on our results. Throughout this paper,  Latin indices refer to tangent space and the Greek indices label coordinates on spacetime. Overcircles quantities are computed from the Levi-Civita connection (Riemannian quantities), we assume the metric signature $(+---)$ and natural units $c=G=1$.

\section{Exact black hole solution in Born-Infeld $f(T)$ gravity }\label{review}
In what follows we provide a brief review on teleparallel gravity and $f(T)$ gravity. These theories are constructed in a manifold which is globally flat and assumes the compatibility condition. Hence, torsion is the field strength tensor of the theory which is the responsible of gravity. Its formulation is usually built from tetrads $e^a{}_\mu$ (orthonormal basis on tangent space) and a purely-gauge spin connection $w^{a}{}_{b\mu}$ (see~\cite{Bahamonde:2021gfp} for more details about these theories). According to this picture, the metric tensor can be expressed in terms of the tetrad field terms of the following equation
\beq
g_{\mu \nu} = \eta_{a b} e^a{}_\mu e^b{}_\nu \ ,  \label{eq:gmunueta}
\eeq
in which $\eta_{a b} = \text{diag}(1,-1,-1,-1)$ is the Minkowski metric. Using them, we define the torsion tensor as
\begin{equation}
T^{\rho}{}_{\mu\nu} = e_a{}^{\rho}\left(\partial_{\mu}e^a{}_{\nu} - \partial_{\nu}e^a{}_{\mu} + \omega^a{}_{b\mu}e^b{}_{\nu} - \omega^a{}_{b\nu}e^b{}_{\mu} \right)\,.\label{eq:deftorsiont}
\end{equation}
It can be shown that the torsion tensor is invariant under local Lorentz transformations if one simultaneously transforms the tetrad and spin connection. Since the spin connection is purely-gauge as $  \omega^a{}_{b\mu} = \Lambda^a{}_c\partial_\mu (\Lambda^{-1})^c{}_b\,$, one can always choose a gauge (known as the Weitzenb\"ock gauge) such that it vanishes.

One of the most remarkable results in these theories is that there is an equivalent description of GR but based purely on torsion, the so-called teleparallel equivalent of GR which is built from the action
\begin{equation}
{\cal{S}} = \frac{1}{16 \pi } \int{ T\, e \, d^4x} + {\cal{S}}_M \ ,
\label{eq: action0}
\end{equation}
with  $e = \text{det} (e^a{}_\mu) = \sqrt{-\text{det}(g_{\mu \nu})}$, ${\cal{S}}_M$ represents the action for the matter fields that it is usually assumed to be coupled minimally to the metric, and $T$ is known as the torsion scalar which is defined as
\beq
  T= \frac{1}{4}T^{\rho\sigma\mu}T_{\rho\sigma \mu} + \frac{1}{2}T^{\mu\sigma\rho}T_{\rho\sigma\mu} - T^\rho{}_{\rho\sigma}T^\mu{}_\mu{}^{\sigma} \,. \label{eq:deftorsions}
\eeq
This teleparallel scalar is connected to the Riemannian Ricci scalar as $\mathring{R}=-T+\tfrac{2}{e}\partial_\mu(eT^{\lambda}{}_{\lambda}{}^{\mu})$. Clearly, \eqref{eq: action0} provides the same equations as the Einstein's field equations since $T$ differs by a boundary term to $\lc{R}$ and appears linearly in the action. One can then modify this action to construct modified teleparallel theories of gravity. One of the most famous ones is the one when one upgrades $T$ by an arbitrary and differentiable function of it, namely
\begin{equation}
{\cal{S}} = \frac{1}{16 \pi } \int{ f(T)\, e \, d^4x} + {\cal{S}}_M \,.
\label{eq: action}
\end{equation}
If we vary this action with respect to $e^a{}_\mu$, one finds that the vacuum field equations in the Weitzenb\"ock gauge are
\begin{eqnarray}\notag
  (\partial_{\mu}f_{T})S_{\nu}{}^{\mu\lambda}+ e^{-1}e^{a}{}_{\nu}\partial_{\mu}(e S_{a}{}^{\mu\lambda})f_{T} - 
  f_{T}T^{\sigma}{}_{\mu \nu}S_{\sigma}{}^{\lambda\mu} - \frac{1}{2}
  f \delta_{\nu}^{\lambda}
  =0\label{fTeq}
\end{eqnarray}
where $f_T=df/dT$ and we defined the superpotential as
\beq
 S_{\rho\mu\nu} = \frac{1}{2}\left(T_{\nu\mu\rho} + T_{\rho\mu\nu} - T_{\mu\nu\rho}\right) - g_{\rho\mu}T^{\sigma}{}_{\sigma\nu} + g_{\rho\nu}T^{\sigma}{}_{\sigma\mu} \,. \label{eq:defcontorsion}
\eeq
The field equations~\eqref{fTeq} have symmetric and antisymmetric contributions. The antisymmetric part of the field equations in spherical symmetry has two branches, one given by a real tetrad and another one by a complex one. 
Recently, two exact black hole solutions in $f(T)$ teleparallel gravity were reported by Bahamonde et al. \cite{Bahamonde:2021srr}. These solutions were found in the second branch with tetrad field being \cite{Bahamonde:2021srr}
\begin{widetext}
\begin{eqnarray}\label{tetrad1}\notag
e_{}^{a}{}_\mu&=&\left(
\begin{array}{cccc}
 0 & i \mathcal{B}(r) & 0 & 0 \\
 i \mathcal{A}(r) \sin\theta \cos\phi & 0 &-  r \sin\phi & -r \sin\theta \cos\theta \cos\phi \\
 i \mathcal{A}(r) \sin\theta \sin\phi & 0 &   r \cos\phi & -r\sin\theta \cos\theta \sin\phi \\
 i \mathcal{A}(r) \cos\theta & 0 & 0 &   r \sin^2\theta \\
\end{array}
\right)\,,
\end{eqnarray}
\end{widetext}
which reproduces the spherically symmetric metric
\begin{eqnarray}
    ds^2&=&\mathcal{A}(r)^2d t^2-\mathcal{B}(r)^2d r^2-r^2d\Omega^2\,,
\end{eqnarray}
where $d\Omega^2=d\theta^2+\sin^2\theta d\phi^2$ is the line element of a two dimensional unit sphere. The torsion tensor for this tetrad is spherically symmetric (and real) and  the torsion scalar behaves as
\begin{eqnarray}
T_{}&=&\frac{2}{r^2 \mathcal{B}^2 \mathcal{A}} \left(\left(\mathcal{B}^2+1\right) \mathcal{A}+2 r \mathcal{A}'\right)\,.\label{eq:torsionscalarcomplex}
\end{eqnarray}

Our main aim is to investigate the phenomenology of this new exact black hole solution. The exact solution is based on a theory motivated from Born-Infeld electromagnetism, which has the following form of the function
\begin{equation}
    f(T)=\lambda\Big(\sqrt{1+\frac{2T}{\lambda}}-1\Big)\,,
\end{equation}
with $\lambda$ being the so-called Born-Infeld parameter. It is interesting to mention that Born-Infeld $f(T)$ gravity was the first modified teleparallel gravity studied in the context of inflation~\cite{Ferraro:2008ey}. If one assumes that $T/\lambda\ll 1$, then $f(T)=T-T^2/(2\lambda)+\mathcal{O}(1/\lambda^2)$ which behaves as TEGR (or GR) and a torsion-squared correction. For this theory and the tetrad~\eqref{tetrad1}, there is the following exact spherically symmetric solution~\cite{Bahamonde:2021srr}
\begin{eqnarray}
    ds^2=\frac{a_1^2 }{r}\Big[\sqrt{\lambda } (a_0 \lambda +r)-2 \tan ^{-1}\left(\frac{\sqrt{\lambda } r}{2}\right)\Big]dt^2- \frac{\lambda ^{5/2} r^5}{(4 + r^2 \lambda)^2}\Big[\sqrt{\lambda } (a_0 \lambda +r)-2 \tan ^{-1}\left(\frac{\sqrt{\lambda } r}{2}\right)\Big]^{-1}dr^2-r^2d\Omega^2\,, \label{eq:com_f_T_metric}
\end{eqnarray}
where $a_0,a_1$ are integration constants. In order to obtain an asymptotically flat spacetime we require that  $a_1^2=1/\sqrt{\lambda}$. Further, to have a smooth transition from the Schwarzschild solution of TEGR ($\lambda\rightarrow \infty$) one needs to set $a_0=-2M/\lambda$. Doing so, and by introducing the parameter 
\begin{eqnarray}
    \alpha=\frac{1}{\lambda}
\end{eqnarray}
we obtain that the metric becomes
\begin{eqnarray}
    ds^2=\Big[1-\frac{2M}{r}-\frac{2 \sqrt{\alpha }}{r} \tan ^{-1}\left(\frac{r}{2 \sqrt{\alpha }}\right)\Big]dt^2- \frac{\alpha ^{-5/2} r^5}{(4 + r^2/\alpha)^2}\Big[1-\frac{2M}{r}-\frac{2 \sqrt{\alpha }}{r} \tan ^{-1}\left(\frac{r}{2 \sqrt{\alpha }}\right)\Big]^{-1}dr^2-r^2d\Omega^2\,.\label{eq:com_f_T_metric}
\end{eqnarray}
Thus, smaller values of $\alpha$ would mean that the deviation from GR would be smaller. Let us compute the ADM mass which coincides with the Komar mass in our case and it is defined as 
\begin{eqnarray}
     \mathcal{M}=\lim_{r \to \infty} \frac{r}{2}\left(1-g^{ab} (\partial_a r)(\partial_b r)   \right)=M+\frac{\pi \sqrt{\alpha}}{2}\,.\label{Admmass}
\end{eqnarray}
One can notice that the ADM mass $\mathcal{M}$ is shifted by $\frac{\pi \sqrt{\alpha}}{2}$ with respect to the mass parameter $M$.

If we now assume that $T/\lambda\ll 1$ and we expand the metric up to $\mathcal{O}(\alpha^2)$, we notice that the metric becomes
\begin{eqnarray}\label{expanmet}
      ds^2=\Big[1-\frac{2 M}{r}+\frac{4\,\alpha}{r^2}-\frac{\pi \sqrt{\alpha} }{ r}\Big]dt^2-\Big[1-\frac{2 M}{r}-\frac{16 M\,\alpha}{ r^3}+\frac{12\alpha}{ r^2}-\frac{\pi \sqrt{\alpha} }{ r}\Big]^{-1}dr^2-r^2d\Omega^2+\mathcal{O}(\alpha^2)\,,
\end{eqnarray}
where one can see that this solution behaves as Schwarzschild in addition to some corrections. It is known that this exact solution contains an event horizon as reported in~\cite{Bahamonde:2021srr}.

\section{Black hole shadows in Born-Infeld $f(T)$ gravity\label{secsha}}
Even though teleparallel gravity expresses the gravitational field through torsion, the geodesic equation still holds for this framework if one assumes that matter is minimally coupled to the metric (and at most to the Levi-Civita connection). This can be seen from the following force-like equation which appears in teleparallel gravity:~\cite{Aldrovandi:2013wha}
\begin{equation}
\frac{d^2x^\mu}{d \delta^2}+{\Gamma}^\mu{}_{\nu \rho}\frac{dx^\nu}{d\delta}\frac{dx^\rho}{d\delta}={K^\mu}_{\nu \rho}\frac{dx^\nu}{d\delta} \frac{dx^\rho}{d\delta}\,,
\label{force_eq}
\end{equation}
\noindent where $\delta$ is the arc-length parameter which parametrizes the worldline of the particles, ${\Gamma}^\mu{}_{\nu \rho}$ is the Weitzenb\"ock connection and $\udt{K}{\lambda}{\mu\nu}=\frac{1}{2}\left(\udt{T}{\lambda}{\mu\nu}+\dut{T}{\mu\nu}{\lambda}+\dut{T}{\nu\mu}{\lambda}\right)$ is the contortion tensor. Since the Levi-Civita connection and the Weitzenb\"ock connection are related as $\lc{\Gamma}^\mu{}_{\nu \rho}={\Gamma}^\mu{}_{\nu \rho}-\udt{K}{\lambda}{\mu\nu}$, one can rewrite the above equation yielding the standard geodesic equation
\begin{equation}
\frac{d^2x^\mu}{d\delta^2}+\lc{\Gamma}^\mu{}_{\nu \rho}\frac{dx^\nu}{d\delta}\frac{dx^\rho}{d \delta}=0\,.
\end{equation}
As it is known, there are two constants of motion  for particle 
motion in spherical symmetry, namely the energy $E$ and the angular momentum 
$L$ of the particle, due to the existence of the timelike 
and spacelike Killing vectors. Following the standard 
procedure it is straightforward to obtain the equations of motion for 
photons as
\begin{eqnarray}
\label{p1a}\frac{dt}{d\delta} &=& \frac{E}{\mathcal{A}^2(r)}\,,\quad
\label{rl}\mathcal{A}(r)\mathcal{B}(r)\frac{dr}{d\delta} = \pm\frac{\sqrt{R(r)}}{r^2}\,,\quad
\frac{d\theta}{d\delta} = \pm\frac{\sqrt{\Theta(\theta)}}{r^2}\,,\quad
\frac{d \phi}{d \delta} = \frac{L \csc^2\theta}{r^2}\,,
\end{eqnarray}
where we have defined
\begin{eqnarray}
R(r) &\equiv&  E^2r^4 - (\mathcal{Q}+L^2) r^2 \mathcal{A}^2(r)\,\,\,\,\text{and}\,\,\,\,\Theta(\theta)=\mathcal{Q}^2-\frac{L}{\sin^2\theta},
\end{eqnarray}
with $\mathcal{Q}$ being the Carter constant. Separation of the Hamilton-Jacobi equations into radial and polar parts involves a seperation constant \cite{Chandra}. Using the above equations we can further study the radial geodesics by 
introducing the effective potential $V_{\text{eff}}(r)$ as follows
\begin{equation}\label{eff}
 \left(\frac{dr}{d\delta}\right)^2 + V_{\text{eff}}(r)= 0\,,
\end{equation}
where
\begin{eqnarray}
 V_{\text{eff}}(r)=-\frac{1}{\mathcal{A}^2(r)\mathcal{B}^2(r)}\frac{R(r)}{r^4}
\end{eqnarray}
and
\begin{equation}
\xi = \frac{L}{E},\;\;\eta = \frac{\mathcal{Q}}{E^2}\,.
\end{equation}

\begin{figure*}[!htb]
 	 	\includegraphics[width=7.5 cm]{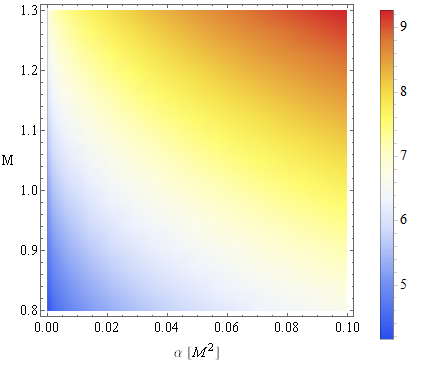}
 	 	 	 	\includegraphics[width=8.5 cm]{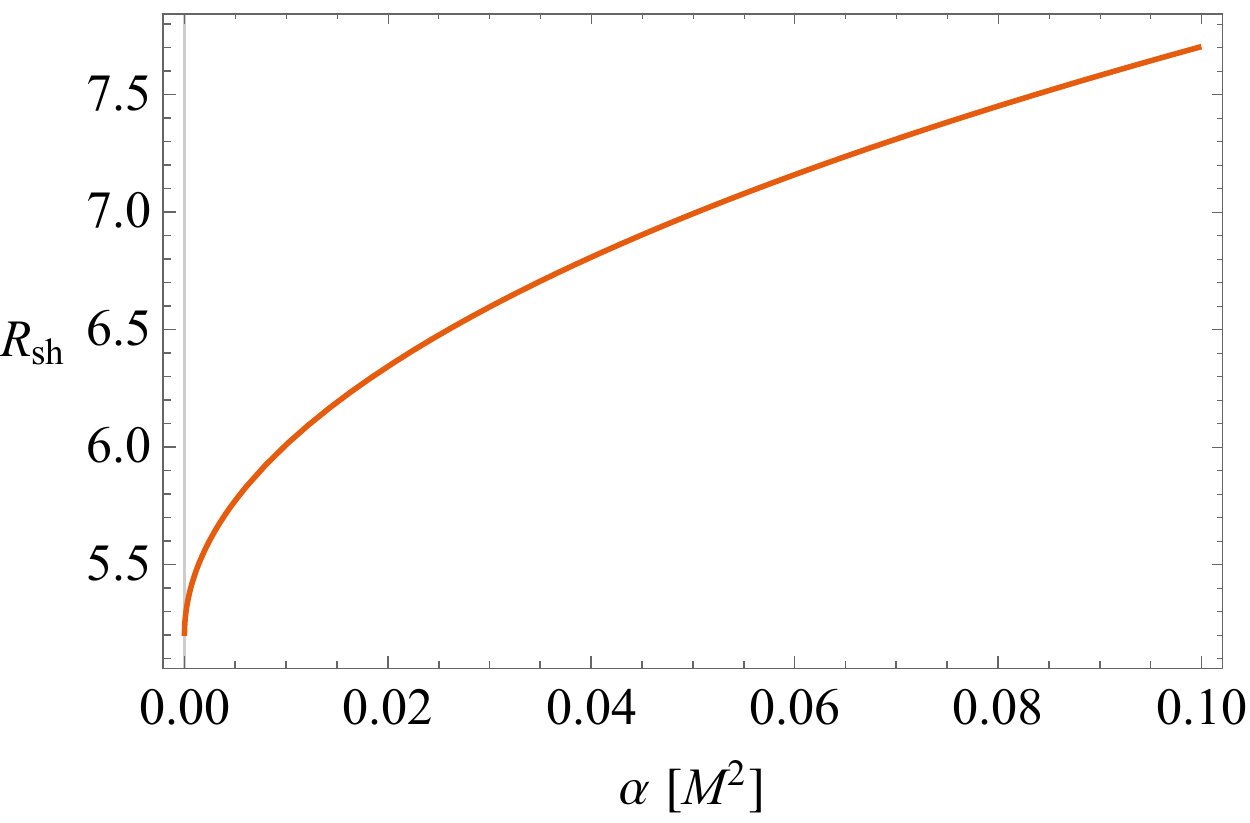}
		\caption{{\it{Left panel: Density plot of the shadow $R_{\rm sh}$ as a function of the parameter $M$ and $\alpha$ for the metric described by~\eqref{eq:com_f_T_metric}. Right panel: Shadow radius as a function of $\alpha$ with a fixed mass $M=1$.
 }}}
\label{Densityplot2}
	\end{figure*}	
	
We can use the two impact parameters $\xi$ and $\eta$ in order to analyze the 
motion of photons around the black hole. Since we are interested to explore the possible deviations from GR encoded in the parameter $\alpha$ we need to use 
the conditions for unstable orbit. As we know, in the observer’s sky, we can observe the black hole shadow due to the fact that some of the scattered photons escape from the black hole while some photons are captured by the black hole. These unstable circular photon orbits can be obtained by applying the following conditions:
\begin{equation}
V_{\text{eff}}(r)=0\,, \quad 
\frac{d V_{\text{eff}}}{dr}=0\,, \quad \frac{d^2V_{\text{eff}}}{dr^2} \leq 0\,.
\end{equation}
Using Eqs.~\eqref{rl} and~\eqref{eff} it is easy to combine $V_{\text{eff}}$ and $R(r)$. If we express the above conditions in terms of $R(r)$ we obtain:
\begin{equation}\lb{condition}
R(r)=0\,,\quad \frac{dR(r)}{dr} =0\,,\quad \frac{d^2 R(r)}{dr^2} >0\,.
\end{equation}

The radius of the photon sphere can be used to find the size of black hole 
shadow. In order to describe the shadow as  seen by large distances,  one introduces 
the two celestial coordinates $X$ and $Y$~\cite{Chandra}, namely
\begin{align}
X &= \lim_{r_* \to \infty} \left(- r_*^2 \sin \theta_0
\frac{d\phi}{dr}\right)\,,\\
Y &= \lim_{r_* \to \infty} r_*^2 \frac{d\theta}{dr}\,,
\end{align}
with $r_*$ the distance between the black hole 
and the observer, and $\theta_0$  the inclination angle 
between the observer's line of sight  and the black 
hole rotational axis.

Using the geodesics equations we finally obtain 
\begin{eqnarray}
 &&X = - \xi(r_{\rm ph}) \csc\theta_0\,, \\
 &&
 Y =  \sqrt{\eta (r_{\rm ph}) -
\xi^2(r_{\rm ph}) \cot^2\theta_0}\,,
\end{eqnarray}
and thus   we have 
$
X^2+Y^2 = \xi^2(r_{\rm ph}) +\eta(r_{\rm ph})$.
Hence, the shadow radius $R_{\rm sh}$ can finally be found 
as
\begin{equation}
R_{\rm sh}(r_{\rm ph}) =\sqrt{\xi^2(r_{\rm ph})+\eta(r_{\rm ph})}=\frac{r_{\rm 
ph}}{ \mathcal{A}(r_{\rm ph})}\,.
\end{equation}
	
The angular diameter of the shadow can be defined as 
\begin{eqnarray}
\theta_{\rm sh}=\frac{2\,R_{\rm sh}M}{D}\,,
\end{eqnarray}
where $D$ is the distance from the black hole and $M$ is the black hole mass. In Fig~\ref{Densityplot2}, we display the shadow radius of the black hole in Born-Infeld $f(T)$ gravity as a function of the parameter $\alpha$. We find that the shadow radius increases with the increase of $\alpha$.

We can see that the event horizon radius is expected to increase due to parameter $\alpha$. We close this section by  considering the scenario where the black hole is surrounded by an infalling/radiating accretion flow. Via this simple model, one  can extract valuable information about the intensity of the radiation which can be detected by a distant observer. In order to achieve this we need to estimate the specific intensity at the observed photon frequency $\nu_\text{obs}$ at the point $(X,Y)$ of the observer's image 
\cite{Narayan:2019imo,Saurabh:2020zqg,Jusufi:2020zln,Zeng:2020dco,Falcke:1999pj,
Bambi:2013nla}
\begin{eqnarray}
I_{\rm obs}(\nu_{\rm obs},X,Y) = \int_{\gamma}\mathrm{g}^3 j(\nu_{e})dl_\text{prop}\,.
\end{eqnarray}
We consider a freely-falling gas dropped from rest at infinity with $E = 1$, which has the following four-velocity
\begin{eqnarray}
u^\mu_e=\Big(\frac{1}{\mathcal{A}^2(r)},-\frac{1}{\mathcal{A}(r) \mathcal{B}(r)}\sqrt{1-\mathcal{A}^2(r)},0,0\Big)\,.
\end{eqnarray}
 In addition we need to use the condition $p_{\mu}p^{\mu}=0$, from which one 
can 
easily obtain
\begin{eqnarray}
    \frac{p^r}{p^t} = \pm \frac{\mathcal{A}^2(r) }{\mathcal{B}(r) }
\sqrt{\bigg(\frac{1}{\mathcal{A}^2(r)}-\frac{b^2}{r^2}\bigg)}\,,
\end{eqnarray}
with $b$ being  the impact parameter. It is 
important to mention here that sign $+(-)$ describes the case when the photon 
approaches (or draws away)
 from the black hole. The redshift function $z$ can be calculated 
using 
\cite{Narayan:2019imo,Saurabh:2020zqg,Jusufi:2020zln,Zeng:2020dco,Falcke:1999pj,
Bambi:2013nla}
\begin{eqnarray}
  z =\frac{p_{\mu}u_{\rm obs}^{\mu}}{p_{\nu}u_e^{\nu}}\,,
\end{eqnarray}
with  $u_{\rm obs}^{\mu}$  the 4-velocity of the 
observer. For the specific emissivity we assume a simple model in which the 
emission is monochromatic, with emitter's-rest frame frequency $\nu_{\star}$, 
and the emission has a $1/r^2$ radial profile:
\begin{eqnarray}
    j(\nu_{e}) \propto \frac{\delta_D(\nu_{e}-\nu_{\star})}{r^2}\,,
\end{eqnarray}
where $\delta_D$ denotes the Dirac delta function. The accretion flow has spherically symmetry and it can be described as an optically-thin disk
region around the black hole.  Expressing the proper length 
in terms of radial coordinate for observed flux, we find
\begin{eqnarray} \label{inten}
    F_{\rm obs}(X,Y) \propto -\int_{\gamma} \frac{z^3\, p_t}{r^2p^r}dr.  
\end{eqnarray}

In the next section we will constrain the parameter $\alpha$ and use its best fit value to plot the shadow images for the Sgr A$^{\star}$ black hole.

\section{Observational constraints on Born-Infeld  $f(T)$ gravity\label{secobv}}
Let us now focus on the most important aspect of the present work which is to constrain the parameter of the theory and test this solution with observations from the shadow of the Sgr A$^\star$ central black hole. For the distance we shall assume the distance $D = 8.3$ kpc, while for the black hole mass we can take $M = 
4.1\times 10^6$ M$\textsubscript{\(\odot\)}.$
Our strategy is to first constrain $\alpha$ using the motion of the S2 star orbit (see, the data \cite{Gillessen:2009ht,GRAVITY:2018ofz}) and then construct the specific shadow images within the range of validity of the observations and more specifically, its best fitting value. To study the motion of a test particle 
using the $f(T)$ solution, without loss of generality, we can take the equatorial plane with $(\theta=\pi/2,~\dot\theta=0)$, and then having in mind that there are two constants of motion (namely the total energy $E$ and the total angular momentum $L$ of the S2 star) which are given by
\begin{equation}
   \dot{t}=\frac{E}{\mathcal{A}^2(r)}\,,\quad \dot{\phi}=\frac{L}{r^2},
\end{equation}
where dots refer differentiation with respect to the arc-length parameter $\delta$. One can also find that the radial equation of motion 
for the particle (S2 star in our case) is given by
\cite{Do:2019txf,Becerra-Vergara:2020xoj,Nampalliwar:2021tyz}
\begin{equation}
     \ddot{r} =-\dfrac{1}{2  \mathcal{B}^2(r)}\left[\frac{d \mathcal{A}^2(r)}{dr} \ \dot{t}^2 - \frac{d \mathcal{B}^2(r)}{dr}
\ \dot{r}^2 -2 r \dot{\phi}^2\right]\,.\label{eqn:motionr}
\end{equation}

The real position is given by the Cartesian coordinates $(x, y, z)$, along with the corresponding velocity components $(v_x, v_y , v_z)$. From the Cartesian coordinates we can find the coordinates in spherical coordinates in the equatorial plane easily using
\begin{align}\label{eqn:xyz}
   x&= r \cos\phi\,,\quad  y= r \sin\phi\,,\quad  z=0\,,
\end{align}
with the corresponding three-velocities given by
\begin{equation}\label{eqn:vxvyvz}
   v_x = v_r \cos\phi - r v_\phi \sin\phi\,,\quad   v_y = v_r \sin\phi + r v_\phi \cos\phi\,,\quad    v_z=0\,.
\end{equation}
During the numerical analyses, from the observers point of view, we need to work with the apparent orbit which has the coordinates  $(\mathcal{X}, \mathcal{Y}, \mathcal{Z})$ related to the real orbit given by $(x, y, z)$ using \cite{13}
\begin{align}
    \mathcal{X} = x B + y G\,,\quad
     \mathcal{Y} = x A+ y F\,,\quad
     \mathcal{Z} = x C + y F\,,
\end{align}
whereas for the apparent components of the velocity and the real component of the velocity we have the relations  \cite{13}
\begin{align}
    \mathcal{V}_X = v_x B + v_y G\,,\quad
     \mathcal{V}_Y = v_x A+ v_y F\,,\quad
     \mathcal{V}_Z = v_x C + v_y F\,,
\end{align}
in which we defined \cite{13}
\begin{eqnarray}
    B &=& \sin\Omega \cos\omega + \cos\Omega \sin\omega \cos i\,, \\
    G &=& -\sin\Omega \sin\omega + \cos\Omega \cos\omega \cos i\,, \\
    A&=& \cos\Omega \cos\omega - \sin\Omega \sin\omega \cos i\,,\\
    F &=& -\cos\Omega \sin\omega - \sin\Omega \cos\omega \cos i\,,\\
    C&=& \sin\omega \sin i\,,\\
    F &=& \cos\omega \sin i\,.
\end{eqnarray}

\begin{figure*}[!htb]
				\includegraphics[width=8.6cm]{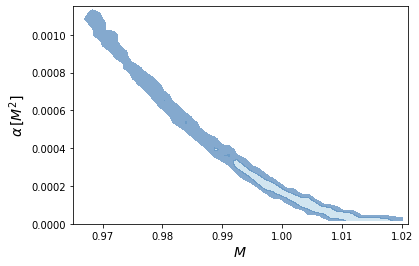}
		\caption{1$\sigma$ and 2$\sigma$  parameter region of Born-Infeld $f(T)$ black hole solution  consistent with  S2 star observations after a Monte-Carlo-Markov Chains analysis.}
		\label{S2stars}
\end{figure*}

\begin{figure*}[!htb]
			\includegraphics[width=8.5 cm]{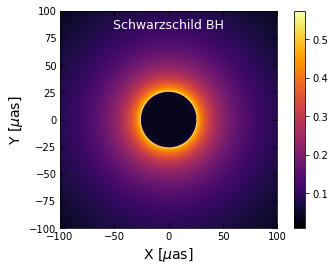}
						\includegraphics[width=8.4 cm]{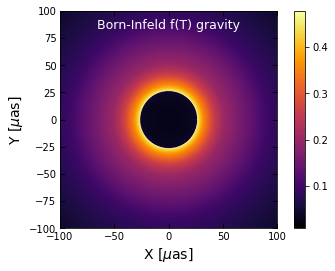}
		\caption{ Shadow images of Sgr A$^\star$ black hole in GR (left panel) and Born-Infeld $f(T)$ gravity (right panel) using the best fit parameters. We have considered the best fit for the black hole mass given in Table I and the distance form the Sgr A$^\star$ to Earth $D=8.3 $ kpc.}\label{shadow}
	\end{figure*}

Note here the following important quantities:  $\omega$, $i$, and $\Omega$ which are the argument of pericenter, the inclination between the real orbit and the observation plane, and the ascending node angle, respectively. Recently, the Gravity Collaboration measured for the first time the orbital precession of S2, and
constrained its departure from the one predicted in GR whose best fit value is $f_{SP}=1.10 \pm 0.19$ \cite{GRAVITY:2020gka}. Here a value $f_{SP}=0$ recovers Newtonian’s gravity and $f_{SP}=1$ which is consistent with GR. The result clearly shows that GR is consistent with observations. In the present work, we would like to test a possible deviations from GR using the modified gravity $f(T)$ gravity solutions. One must relay on the  numerical analyses to obtain the equations of motion for the orbit. We are going to use  the Bayesian theorem with the likelihood 
function with the posterior probability density with the likelihood function is given by 
\begin{eqnarray}
    \ln \mathcal{L}(O|P)=-\frac{1}{2}\sum_{i=1}^{N}\left[ \frac{\left(X_{\textrm{ obs},i}-X_{\textrm{mod},i}\right)^2}{\sigma_{\textrm{obs},i}^2}\right]- \frac{1}{2}\sum_{i=1}^{N}\left[\frac{\left(Y_{\textrm{obs},i}-Y_{\textrm{mod},i}\right)^2}{\sigma_{\textrm{obs},i}^2}\right]\,,
\end{eqnarray}
where the two observed and theoretical quantities are noted as $(X_{\rm obs},Y_{\rm obs})$, and $(X_{mod},Y_{mod})$ (see, \cite{GRAVITY:2018ofz,Do:2019txf}). In order to find the best-fit values we use the Monte-Carlo-Markov Chains analysis. For the 
central mass object we take $4.1 \times 10^6 M_{\odot}$ along with the uniform 
priors $\alpha/M^2 \in [0,1]$ for the Solution 2, respectively. 
In Fig.~\ref{S2stars}, we present 
 the region of the parameter space in agreement with S2 star data. 
Concerning  the $\alpha$ parameter, in which we are interested in this manuscript, the best fit  within 1$\sigma$ confidence are given in Table I. 
\begin{table}[ht]
\begin{center}
\begin{tabular}{|l|l|l|}
\hline
   \,\,\, & 
\,\rm{Schwarzschild (GR)} & \,\rm{Born-Infeld} $f(T)$ gravity \\ \hline
$M\, [ 4.1 \times 10^6 M_\odot]$  & $1.020$ & $1.015$ \\ 
$\alpha\,\,\,\, [M^2]$   & \ \ \ \ \ -   & $1.559 \times 10^{-5}$\\ 
$\Omega\,\,\,\,[^{0}$]   & $228.171$ & 228.171\\ 
$\omega\,\,\,\,[^{0}$]    & $66.263$ & 66.263\\ 
$i\,\,\,\,\,\,[^{0}$]  &  $134.567$ & $134.567$\\ 
$\epsilon$\,\,\,\,\,\,[no units]   & $0.890$ & $ 0.890$\\ 
$a\,\,\,\,\,\,$[mas]  &  $126.332$ & $126.321$\\ 
$\theta_{\rm sh}\,\,[\mu$as]  & $51.686$  & $52.530$ \\ 
$\chi^2_r$   & $6.777$&  $8.801$\\ 
$<\chi^2> $  & $3.885$ & $4.400$\\ 
 \hline
\end{tabular}
{\caption[]{In the Table we present the best fit parameters obtained within $1\sigma$ confidence. }}
\end{center}
\label{TableI}
\end{table}
Within $1\sigma$ we found an upper bound $0\leq \alpha/M^2 \leq 6 \times 10^{-4}$ for the Born-Infeld $f(T)$ gravity. Using the constraints we have presented the shadow of Sgr A$^{\star}$ black hole as shown in Fig.~\ref{shadow}. For the angular diameter we have found $52.530 \,\mu$arcsec. Using the recent result reported for Sgr A$^{\star}$ given by $(51.8 \pm 2.3) \mu$arcsec \cite{EHT2022-1,EHT2022-2,EHT2022-3} we conclude that the angular diameter of the Born-Infeld $f(T)$ gravity is consistent with the observation and could not be distinguished from the Schwarzschild black hole in GR by the present technology. From the shadow images given in Fig.~\ref{shadow}, one can observe the characteristic photon rings and also an interesting fact that the intensities of the electromagnetic radiation are small in Born-Infeld $f(T)$ gravity models compared to Schwarzschild black hole. 

\section{Deflection of Light in Born-Infeld $f(T)$ gravity}\label{def}
    In this final section, we will proceed to elaborate the gravitational lensing effect in weak gravity regime in $f(T)$ gravity models using the optical geometry. For the general case, in the equatorial plane by letting $ds^2=0$, we find the optical metric 
    \begin{eqnarray}
        dt^2=\frac{\mathcal{B}^2(r) dr^2}{\mathcal{A}^2(r)}+\frac{r^2 d\phi^2}{\mathcal{A}^2(r)}.
    \end{eqnarray}
    
To simplify further the work, we can write the optical metric in approximated form in leading order in $\alpha$, in the equatorial plane, given by (see the expanded metric in~\eqref{expanmet})
\begin{equation}
dt^{2}\simeq \frac{dr^2}{\left[1-\frac{2 M}{r}+\frac{4\,\alpha}{r^2}-\frac{\pi \sqrt{\alpha} }{ r}\right] \left[  1-\frac{2 M}{r}+\frac{12\,\alpha}{r^2}-\frac{\pi \sqrt{\alpha} }{ r}\right]}+\frac{r^2 d\phi^2}{1-\frac{2 M}{r}+\frac{4\,\alpha}{r^2}-\frac{\pi \sqrt{\alpha} }{ r}}+\mathcal{O}(\alpha^2)
\end{equation}

In what follows we shall apply the optical geometry by applying the Gauss-Bonnet theorem to compute the deflection angle of light, which can be summarised as follows:\\ 
\textbf{Theorem} \textit{Let $\mathcal{D}_{R}$ be a non-singular domain outside the light ray with boundaries $\partial 
\mathcal{D}_{R}=\gamma_{g^{(\rm op)}}\cup C_{R}$, of an oriented two-dimensional surface $S$ with the optical metric $g^{(\rm op)}$. Moreover let $\mathcal{K}$ and $\kappa $ be the Gaussian optical curvature and the geodesic curvature associated to the optical geometry. Under this construction the Gauss-Bonnet theorem can be written as follows} \cite{Gibbons}
\begin{equation}
\int\limits_{\mathcal{D}_{R}}\mathcal{K}\,dS+\oint\limits_{\partial \mathcal{
A}_{R}}\kappa \,dt+\sum_{k}\theta _{k}=2\pi \chi (\mathcal{D}_{R})\,.
\label{10}
\end{equation}

In this equation $\theta _{k}$ stands for the exterior angle at the $k^{\rm th}$ vertex. We chose the regular domain to be outside of the light ray, say in the $(r,\phi)$ optical plane, this means that the domain has the topology of disc having the  Euler characteristic number $\chi (\mathcal{D}_{R})=1$. Furthermore we need to compute two important quantities: the Gaussian optical curvature and the geodesic curvature. Let us first focus on the Gaussian optical curvature $\mathcal{K}$ which in leading order terms gives 
\begin{eqnarray}
\mathcal{K} & \simeq & - \frac{2M}{r^3}-\frac{\pi \sqrt{\alpha}}{r^3}+\mathcal{O}(\alpha^2)\,.
\end{eqnarray}

Let us now introduce a smooth curve defined as follows $\gamma:=\{t\}\to \mathcal{D}_{R}$,  where the geodesic curvature can be found from \cite{Gibbons}
\begin{equation}
\kappa =g^{(\rm op)}\,\left( \lc{\nabla} _{\dot{\gamma}}\dot{\gamma},\ddot{\gamma}
\right),  
\end{equation}
having in mind the additional unit speed condition $g^{(\rm op)}(\dot{\gamma},\dot{\gamma})=1$, along with the $\ddot{\gamma}$ being the unit acceleration vector. In the large limit, but finite radial distance $r \equiv R\rightarrow \infty $, we have two jump angles (say at the source $\mathcal{S}$ and observer $\mathcal{O})
$, which satisfy  $\theta _{\mathit{O}}+\theta _{\mathit{S}}\rightarrow \pi $ \cite{Gibbons}. On the other hand, for the geodesic curvature for the light ray (geodesics) we must have $\kappa (\gamma_{g^{(\rm op)}})=0$. Hence we only need to compute the contribution to the curve $C_{R}$. From the Gauss-Bonnet theorem we have
\begin{equation}
\lim_{R\rightarrow \infty }\int_{0}^{\pi+\hat{\alpha}_{\rm DA}}\left[\kappa \frac{d t}{d \phi}\right]_{C_R} d \phi=\pi-\lim_{R\rightarrow \infty }\iint\limits_{\mathcal{D}_{R}}\mathcal{K} \,dS.
\end{equation}
To simplify the problem further, let us chose the black hole at the coodinate center and consequently the geodesic curvature for the curve is $C_{R}$ at the distance distance $R$ from the coordinate center. It follows that  \cite{Gibbons}
\begin{equation}
\kappa (C_{R})=|\lc{\nabla} _{\dot{C}_{R}}\dot{C}_{R}|.
\end{equation}
We can now utilize the unit speed condition and the optical metric to show that$
\lim_{R\rightarrow \infty }\kappa (C_{R}) \rightarrow  R^{-1}$. This shows that the optical metric is asymptotically Euclidean and therefore the deflection angle reads
\begin{equation}
\hat{\alpha}_{\rm DA}=-\int\limits_{0}^{\pi }\int\limits_{r(\phi)
}^{\infty } \mathcal{K} dS,
\end{equation}
where we have the surface element reads $dS =\sqrt{g^{(\rm op)}}dr d\phi$ and the equation for the light ray $r(\phi)=b/\sin \phi$. Solving this integral we can approximate the deflection angle as
\begin{equation}
\hat{\alpha}_{\rm DA}= -\int\limits_{0}^{\pi }\int\limits_{\frac{\mathsf{b}}{\sin \phi }
}^{\infty }\left[- \frac{2M}{r^3}-\frac{\pi \sqrt{\alpha}}{r^3}\right]dS=\frac{4M}{b}+\frac{2 \pi \sqrt{\alpha}}{b}-\frac{\pi \left( 3 \pi^2+80  \right) \alpha}{16\, b^2}+\mathcal{O}(\alpha^2)\,.
\end{equation}
From the last equation, one notices that the deflection angle is affected by the parameter $\alpha$ in leading order terms. At least mathematically, in leading order terms, it seems like we cannot distinguish the black hole in $f(T)$ gravity from the GR. Implying that the deflection angle in leading order terms expressed by the ADM mass (see Eq.~\eqref{Admmass}) is given by 
\begin{eqnarray}
    \hat{\alpha}_{\rm DA}=\frac{4}{b}\left(M+\frac{\pi \sqrt{\alpha}}{2}\right)-\frac{\pi \left( 3 \pi^2+80  \right) \alpha}{16\, b^2}=\frac{4 \mathcal{M}}{b}-\frac{\pi \left( 3 \pi^2+80  \right) \alpha}{16\, b^2}+\mathcal{O}(\alpha^2)\,.
\end{eqnarray}
However from the physical point of view the ADM mass in $f(T)$ gravity, i.e. $ \mathcal{M}$ is bigger compared to the mass parameter $M$, since $\alpha\geq 0$.

Of course, we see that one can go beyond the leading order term and, in that case, the second order term is a correction term proportional to $\delta  \hat{\alpha}\sim -\alpha/b^2$, meaning that the deflection angle decreases to the $\alpha$ corrections, but the first order term which increases the mass is dominant and overall the deflection angle is expected to increase.  This increase on the deflection angle explains why the intensity of electromagnetic radiation detected by a distant observer shown in Fig.~\ref{shadow} is smaller in $f(T)$ gravity compared to GR.  The best accuracy of measuring the deflection of light by the sun is from measuring the deflection of radio waves from distant quasars using the Very Long Baseline Array (VLBA)
\cite{vlbi}, which achieved an accuracy of $3 \times 10^{-4}$. For that we can use the leading order contribution term
\begin{eqnarray}
    \frac{\delta \hat{\alpha}_{\rm DA} }{\hat{\alpha}^{GR}_{\rm DA}}=\frac{2 \pi \sqrt{\alpha} /b}{4M/b}< 3 \times 10^{-4}.
\end{eqnarray}
Assuming that light grazes the surface of the Sun  with  $b=6.96 \times  10^8 $ m and mass in geometric units $M=1.48$ km, we get the following interval $0\leq \alpha/M^2 \leq 3.64 \times 10^{-8}$. Where we have excluded the negative domain, we can see that the positive bound belongs to the interval obtained from the S2 star constrain given by  $0\leq \alpha/M^2 \leq 6 \times 10^{-4}$. 

\section{Conclusions}\label{seccon}
We have used an exact solution in Born-Infeld $f(T)$ gravity and studied the phenomenological aspects using the black hole shadows and the S2 star orbit. The impact of modified gravity is encoded in the coupling parameter $\alpha$ which changes the properties of the black hole and 
its shadow. We used data from the orbital motion of 
S2 star around the Sgr A$^\star$ black hole through a Monte-Carlo-Markov Chains (MCMC) analysis within $1\sigma$ we found an upper bound $0\leq \alpha/M^2<6 \times 10^{-4}$. Using the best fit parameters we found the angular diameter $52.530\, \mu$arcsec for the shadow.  Using the recent result for the angular diameter for Sgr A$^{\star}$ reported as $(51.8 \pm 2.3 ) \mu$arcsec \cite{EHT2022-1,EHT2022-2,EHT2022-3} we see that the angular diameter of the Born-Infeld $f(T)$ gravity is consistent with the observation and very difficult to distinguished from the Schwarzschild black hole, at least with the present data. We have also computed the deflection angle of light in the weak gravity regime and found that in leading order terms mathematically the  Born-Infeld $f(T)$ gravity cannot be distinguished from the classical GR, however the ADM mass in $f(T)$ gravity is bigger due to $\alpha$ corrections and overall the deflection angle is expected to increase. This fact is in perfect agreement with our shadow image shown in Fig.~\ref{shadow}, for Born-Infeld $f(T)$ gravity the intensity observed at infinity is smaller compared to GR. This has to do with the fact that when the deflection angle increases more photons will be captured by the black hole and the intensity will be smaller at infinity. From the weak deflection angle we obtain an upper bound given by $0\leq \alpha/M^2 \leq 3.64 \times 10^{-8}$. Finally, it should be noted that, with the improvement of the precision of measuring the shadow of black holes we should have better constraints to have a more rigorous test of modified theories of gravity. 

Still there are some questions regarding the exact black hole solution that we analysed. For example, it is important to analyse its stability by performing perturbations of the tetrad. Furthermore, some other aspects that could be interesting to analyse are the predictions of the quasi-normal modes generated by having this background solution. Since the tetrad is complex, one would need to be careful on performing these perturbations carefully such that all the observables are real. Another question that could be interesting astrophysically is to analyse its viability by studying accretion disks configurations under this solution. All of these studies will be reported in forthcoming studies.

\subsection*{Acknowledgements}
SC acknowledges the support of Istituto Nazionale di Fisica Nucleare (INFN), {\it iniziative specifiche}  QGSKY and Moonlight2. S.B. is supported by JSPS Postdoctoral Fellowships for Research in Japan and KAKENHI Grant-in-Aid for Scientific Research No. JP21F21789. S.B. also acknowledges the Estonian Research Council grants PRG356 ``Gauge Gravity"  and the European Regional Development Fund through the Center of Excellence TK133 ``The Dark Side of the Universe". We would like to thank the referee for the careful revision and the constructive comments.

\end{document}